\documentclass[twocolumn,showpacs,preprintnumbers,superscriptaddress,amsmath,%
amssymb,nofootinbib]{revtex4}

\input psfig.sty

\usepackage{graphicx}% Include figure files
\usepackage{dcolumn}% Align table columns on decimal point
\usepackage{bm}% bold math

\newcommand{\be}{\begin{equation}}
\newcommand{\ee}{\end{equation}}

\begin{document}
\title{Energy Conditions and Cosmic Acceleration}

\author{J. Santos} \email{janilo@on.br}

\affiliation{Universidade Federal do Rio Grande do Norte, Departamento de
F\'{\i}sica, C.P. 1641, 59072-970 Natal -- RN, Brasil}

\affiliation{Departamento de Astronomia, Observat\'orio Nacional, 
%Rua Gal.\ Jos\'e Cristino 77 \\
20921-400, Rio de Janeiro -- RJ, Brasil}

\affiliation{Centro Brasileiro de Pesquisas F\'{\i}sicas, 
%Rua Dr.\ Xavier Sigaud 150 \\
22290-180, Rio de Janeiro -- RJ, Brasil}

\author{J.S. Alcaniz}\email{alcaniz@on.br}
\affiliation{Departamento de Astronomia, Observat\'orio Nacional, 
%Rua Gal.\ Jos\'e Cristino 77 \\
20921-400, Rio de Janeiro -- RJ, Brasil}

\author{N. Pires} \email{npires@dfte.ufrn.br}

\affiliation{Universidade Federal do Rio Grande do Norte, Departamento de
F\'{\i}sica, C.P. 1641, 59072-970 Natal -- RN, Brasil}

\author{M.J. Rebou\c{c}as}\email{reboucas@cbpf.br}
\affiliation{Centro Brasileiro de Pesquisas F\'{\i}sicas, 
%Rua Dr.\ Xavier Sigaud 150 \\
22290-180, Rio de Janeiro -- RJ, Brasil}

\date{\today}

\begin{abstract}
In general relativity, the energy conditions are invoked
to restrict general energy-momentum tensors $T_{\mu\nu}$
in different frameworks, and to derive general results that hold
in a variety of general contexts on physical grounds.
We show that in the standard Friedmann--Lema\^{\i}tre--Robertson--Walker
(FLRW) approach,  where the equation of state of the cosmological fluid 
is unknown, the energy conditions provide model-independent bounds on 
the behavior of the distance modulus of cosmic sources as a function 
of the redshift for any spatial curvature. We use the most recent type 
Ia supernovae (SNe Ia) observations, which include the new \emph{Hubble 
Space Telescope} SNe Ia events,  to carry out a model-independent analysis 
of the energy conditions violation in the context of the standard cosmology. 
We show that both the null (NEC), weak (WEC) and dominant (DEC) conditions, which are associated with the existence of the so-called \emph{phantom} fields, seem to have 
been violated 
only recently ($z \lesssim 0.2$), whereas the condition for attractive gravity, 
i.e., the strong energy condition (SEC) was firstly violated billions of 
years ago, at $z \gtrsim 1$.
\end{abstract}

%\pacs{98.80.Jk; 98.80.-k; 04.20.Cv}

\maketitle

%%%%%%%%%%%%%%%%%%%%%%%%%%%
\section{Introduction}
%%%%%%%%%%%%%%%%%%%%%%%%%%%

Within the framework of the standard Friedmann--Lema\^{\i}tre--Robertson%
--Walker (FLRW) cosmology, the Universe is modelled by a space-time manifold
endowed with a spatially homogeneous and isotropic  metric
\begin{equation}
\label{RWmetric} ds^2 = dt^2 - a^2 (t) \left[\, \frac{dr^2}{1-kr^2} +
r^2(d\theta^2 + \sin^2 \theta  d\phi^2) \,\right]\,,
\end{equation}
where the spatial curvature $k=0,\, 1,\, \mbox{or} -1$, $a(t)$ is the  scale
factor, and we have set the speed of light $c = 1$. The metric~(\ref{RWmetric})
only expresses the principle of spatial homogeneity and isotropy along with the
existence of a cosmic time $t$. However, to study the dynamics of the Universe
a third assumption in this approach to cosmological modelling is necessary,
namely, that the large scale structure of the Universe is essentially determined by
gravitational interactions, and hence can be described by a  metrical theory
of gravitation such as  general relativity (GR), which we assume in this work.

These very general assumptions constrain the cosmological fluid to
be a perfect-type fluid
\begin{equation} \label{EM-Tensor}
T_{\mu\nu} = (\rho+p)\,u_\mu u_\nu - p \,g_{\mu \nu}\;,
\end{equation}
where $u_\mu$ is the fluid four-velocity, with total density $\rho$ and
pressure $p$ given, respectively, by 
\begin{eqnarray}
\rho & = & \frac{3}{8\pi G}\left[\,\frac{\dot{a}^2}{a^2}
                                        +\frac{k}{a^2} \,\right]\;,
\label{rho-eq} \\
p & = & - \frac{1}{8\pi G}\left[\, 2\,\frac{\ddot{a}}{a} +
\frac{\dot{a}^2}{a^2} + \frac{k}{a^2} \,\right] \;, \label{p-eq}
\end{eqnarray}
where $G$ is the Newton constant, and dots indicate derivative
with respect to the time $t$.

A further constraint on this standard cosmological picture, without
invoking any particular equation of state, arises from the
so-called \emph{energy conditions\/}~\cite{Hawking-Ellis,Visser,Carroll}
that limit the  energy-momentum tensor $T_{\mu\nu}$ on physical grounds.
These conditions can be stated in a coordinate-invariant way,
in terms of $T_{\mu\nu}$ and vector fields of fixed character
(timelike, null and spacelike). In the FLRW framework, however,
only the energy-momentum of a perfect fluid~(\ref{EM-Tensor})
should be considered, so that the most common energy
conditions (see, e.g.,
\cite{Hawking-Ellis,Visser,Carroll,Visser-Barcelo})
reduce to
%%%
\[
\begin{array}{llll}
\mbox{\bf NEC} \ & \Longrightarrow &\, \rho + p \geq 0 \;,  &   \\
\\
\mbox{\bf WEC} \ & \Longrightarrow & \rho \geq 0 &
\ \mbox{and} \quad\, \rho + p \geq 0 \;,  \\
\\
\mbox{\bf SEC}   &\Longrightarrow & \rho + 3p \geq 0 &
\ \mbox{and} \quad\, \rho + p \geq 0 \;, \\
\\
\mbox{\bf DEC}   &\Longrightarrow & \rho \geq 0  &
\ \mbox{and} \; -\rho \leq p \leq\rho \;,
\end{array}
\]
where NEC, WEC, SEC and DEC correspond, respectively, to the null, 
weak, strong and dominant energy conditions.
From  Eqs.~(\ref{rho-eq})~--~(\ref{p-eq}), one easily obtains
that these energy conditions can be translated into the following set 
of dynamical constraints relating the scale factor $a(t)$ and its 
derivatives for any spatial curvature $k\,$:
\begin{eqnarray}
\mbox{\bf NEC} & \, \Longrightarrow & \quad\, - \frac{\ddot{a}}{a}
+  \frac{\dot{a}^2}{a^2}  + \frac{k}{a^2} \geq 0 \;, \label{nec-eq}
\\
\mbox{\bf WEC} & \, \Longrightarrow & \quad\;\; \frac{\dot{a}^2}{a^ 2} 
+ \frac{k}{a^2} \geq 0 \;,
\label{wec-eq}
\\
\mbox{\bf SEC} & \, \Longrightarrow & \quad\;\; \frac{\ddot{a}}{a} \leq 0 \;,
\label{sec-eq} \\
\mbox{\bf DEC} & \, \Longrightarrow & \; - 2\left[
\frac{\dot{a}^2}{a^2}+\frac{k}{a^2} \right] \leq
\frac{\ddot{a}}{a}  \leq
\frac{\dot{a}^2}{a^2}+\frac{k}{a^2} \;. \label{dec-eq}
\end{eqnarray}
where clearly the NEC restriction [Eq.~(\ref{nec-eq})] is also part of
the WEC constraints. 
{}From the theoretical viewpoint, these energy conditions have been 
used in different contexts to derive general results that hold for 
a variety of situations~\cite{ec1}. 
For example, the Hawking-Penrose singularity theorems invoke the WEC 
and SEC~\cite{Hawking-Ellis}, whereas the proof of the second law of
black hole thermodynamics requires the NEC \cite{Visser}.

\begin{figure*}[t]
%\vspace{.2in}
\centerline{\psfig{figure=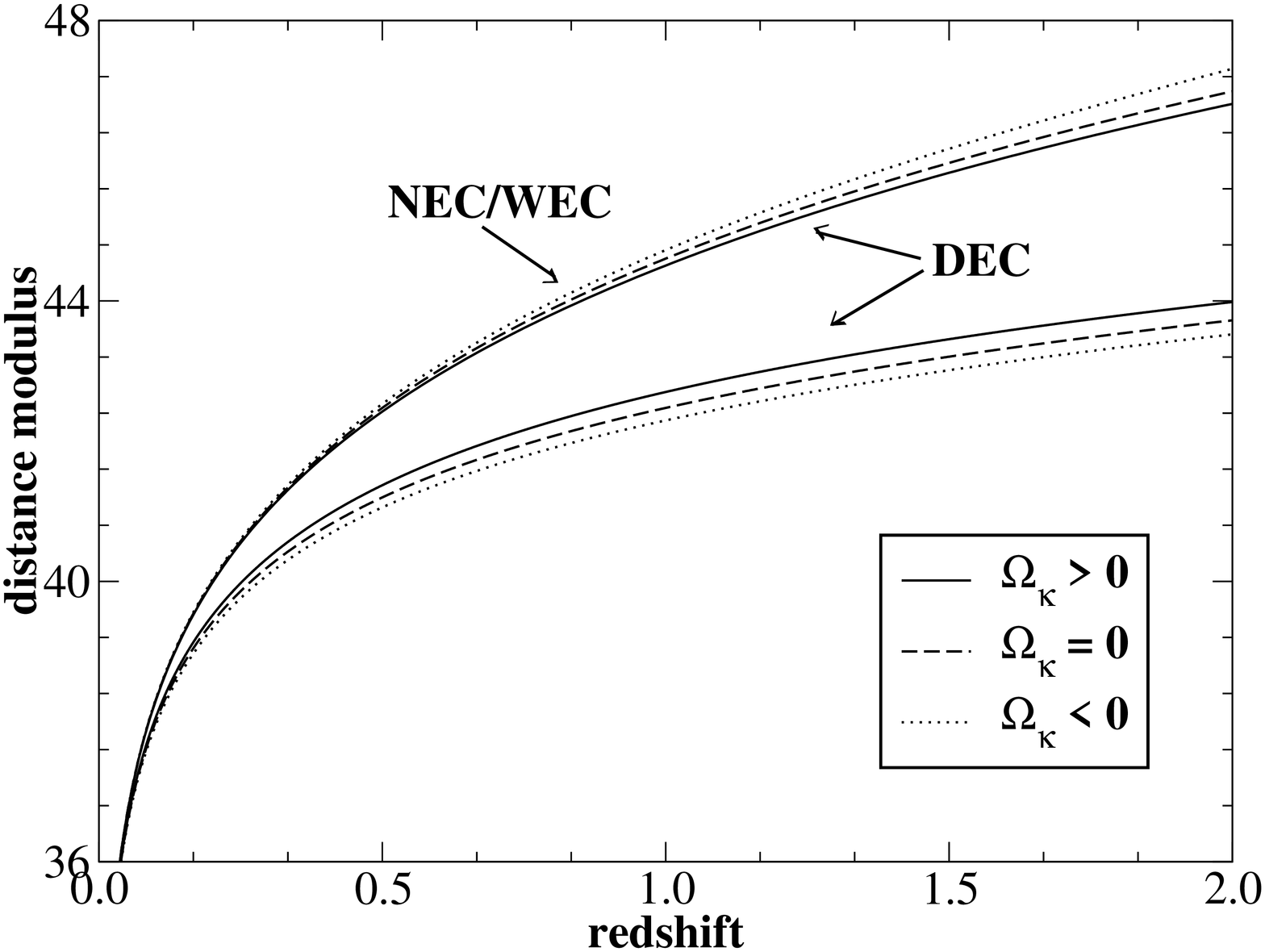,width=3.4truein,height=3.2truein,angle=-90} 
\psfig{figure=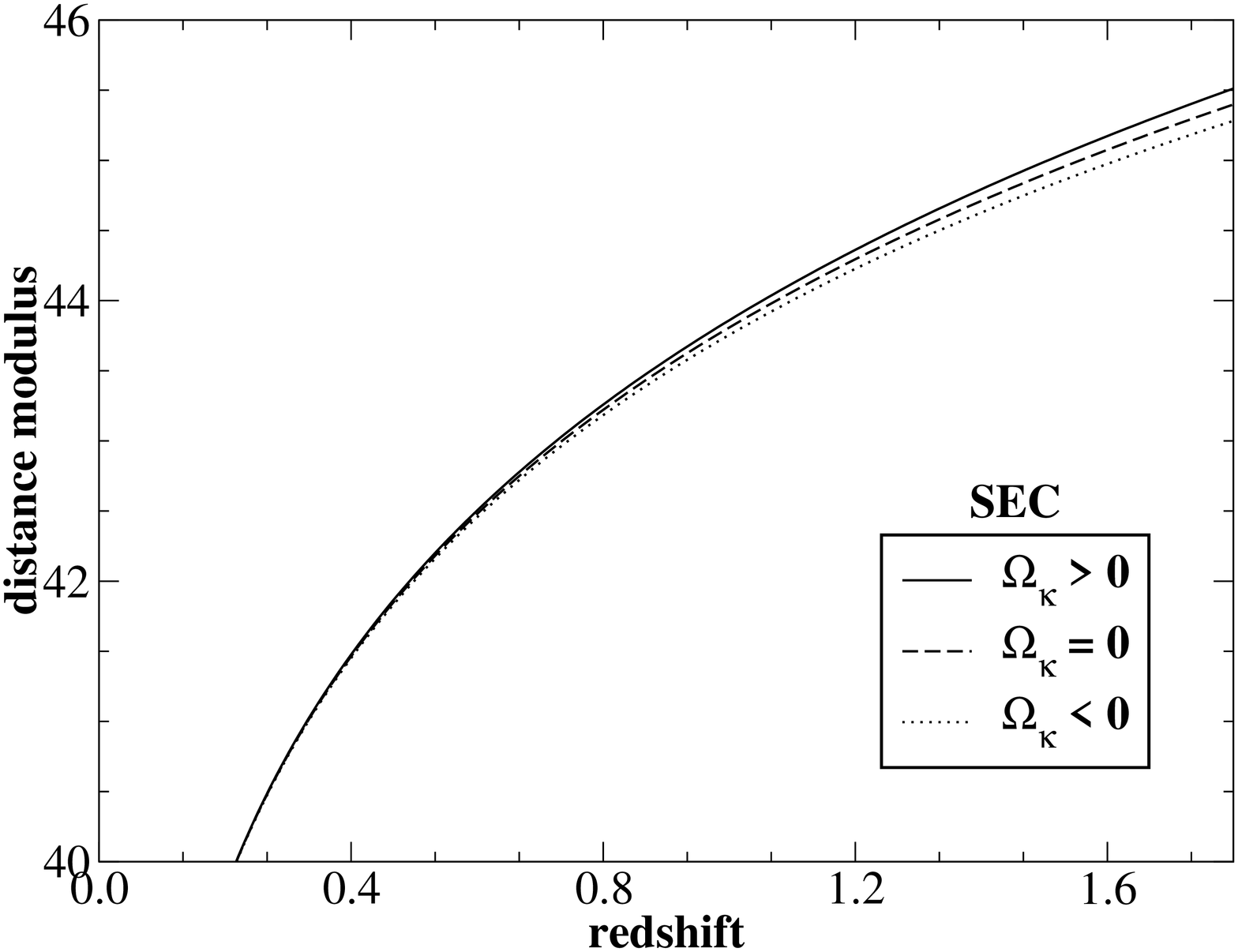,width=3.4truein,height=3.2truein,angle=-90}
\hskip 0.3in} 
\caption{Model-independent bounds on the distance modulus $\mu(z)$ as a 
function of the redshift for different signs of the curvature parameter $\Omega_k$.\ 
Panel {\bf (1a)}: Bounds from the NEC/WEC are shown in the top set of curves, 
while the upper and lower bounds from the DEC correspond, respectively, 
to the top and bottom sets of curves. Panel {\bf (1b)}:  Upper bounds from the SEC for all signs of the spatial curvature.} 
\end{figure*}

In order to shed some light on the energy conditions interrelations 
from the observational side,  it is important to confront the constraints 
arising from Eqs.~(\ref{nec-eq})~--~(\ref{dec-eq}) with the current 
observational data. In this regard, recently some of us~\cite{SAR2006} 
used the fact that the classical energy conditions can be translated 
into differential constraints involving only the scale factor and its
derivatives,  to place model-independent bounds on the distance modulus 
$\mu(z)$ of cosmic sources in a \emph{flat} ($k=0$) FLRW universe. 
When compared with the Type Ia Supernovae (SNe Ia) data, as compiled 
by Riess \emph{et al.\/}~\cite{Riess2004}, and Astier 
\emph{et al.\/}~\cite{Legacy2005}, it was shown that all the
energy conditions seem to have been violated in a recent past of
the cosmic evolution ($z \sim 1$), when the Universe is expected to
be dominated by \emph{normal} matter fields.

The aim of this paper is twofold. First, we extend the results of 
Ref.~\cite{SAR2006} by deriving model-independent bounds on $\mu(z)$ 
for any spatial curvature $k$, including the flat one ($k=0$) 
as a special limiting case.  Second, we confront our general bounds
with the most recent SNe Ia observations, as provided recently
by  Riess \emph{et al.}~\cite{Riess2006}, which include the new 
\emph{Hubble Space Telescope\/} (HST) SNe Ia events. This new data 
sample indicates that both NEC and DEC were violated only recently 
($z \lesssim 0.2$), whereas the condition for attractive gravity (SEC) 
was firstly violated billions of years ago, at $z \gtrsim 1$.

%%%%%%%%%%%%%%%%%%%%%%%%%%%%%%%%%%%%%%%%%%%%%%%%%%%%%%%%%%%%%%%%%
\section{General Constraints on Distances from Energy Conditions}
%%%%%%%%%%%%%%%%%%%%%%%%%%%%%%%%%%%%%%%%%%%%%%%%%%%%%%%%%%%%%%%%%

The predicted distance modulus for an object at redshift $z$ is 
defined as
\begin{equation} \label{dist-mod}
\mu(z) \equiv m(z) - M = 5\,\log_{10} d_L (z) + 25\;,
\end{equation}
where $m$ and $M$ are, respectively, the apparent and absolute 
magnitudes, and $d_L$, given by 
\begin{equation}  \label{dist-lumin}
d_L(z) = a_0\,(1+z)\,r(a) \;,
\end{equation}
stands for the luminosity distance in units of megaparsecs 
(throughout this paper the subscript $0$ denotes present-day 
quantities). From Eq.~(1), it is straightforward to show that 
the radial distance $r(a)$ can be written as
\begin{equation}   \label{proper-distance}
r(a)=\frac{H_0^{-1}}{a_0\sqrt{|\Omega_k|}}\,\,S_{k}\left[
\sqrt{|\Omega_k|}\,I(a)\right]\;,
\end{equation}
where $\Omega_k = -k/(a_0H_0)^2$ is the usual definition of \vspace{2mm} 
the curvature parameter, $I(a)$ is given by 
\begin{equation}  \label{I(a)} 
I(a)= a_0\,H_0\int_a^{a_0} \frac{da}{\dot{a}a}\;,
\end{equation} 
and the function $S_{k}(x)$ takes one of the following 
forms:
\begin{eqnarray} \label{S-function}
S_{k}(x)\equiv \left\{
\begin{array}{ll}
\sin(x) & \mbox{if $\quad\Omega_k < 0$} \;,\\
x & \mbox{if $\quad\Omega_k = 0$} \;,\\
\sinh(x) & \mbox{if $\quad\Omega_k > 0$ \;.}
\end{array} \right.
\\ \nonumber
\end{eqnarray}

%%%%%%%%%%%%%%%%%%%%%%%%%%
\vspace{0.3cm}
\centerline{\bf{A.  NEC/WEC}}
\vspace{0.3cm}
%%%%%%%%%%%%%%%%%%%%%%%%%%

In order to derive bounds on the predicted distance modulus $\mu(z)$ 
from NEC/WEC we note that the first integral of Eq.~(\ref{nec-eq}) 
provides
\begin{equation}
\dot{a} \geq a_0\,H_0 \,\sqrt{\Omega_k+(1-\Omega_k)(a/a_0)^2}\;,
\end{equation}
for any value of $a < a_0$. By using the above inequality we 
integrate~(\ref{I(a)}) to obtain the following upper bound on 
the radial distance:
\begin{widetext}
\begin{eqnarray}
r(z) & \leq & \frac{H_0^{-1}}{ a_0\sqrt{|\Omega_k|}} %\,\,
S_{k}\left\{ S^{-1}_{k} \left[
\sqrt{\left|\frac{\Omega_k}{\Omega_k-1}\right|}\,
(1+z)\right] \right. %\nonumber \\
%& & 
\left. - S^{-1}_{k}
\sqrt{\left|\frac{\Omega_k}{\Omega_k-1}\right|}\, \right\}\;,
  \label{comoving-distance-WEC}
\end{eqnarray}
\end{widetext}
where $a_0/a = 1 + z\,$, and $S_{k}^{-1}$ is the inverse function 
of $S_{k}$.  Concerning the derivation of the above expression, 
two important aspects should be emphasized at this point. First, that it uses 
the constraint $\Omega_k<1$ that arises from the WEC, as given by Eq.~(\ref{wec-eq}).
Second,  since the argument of the function $\sin^{-1}(x)$ is limited to 
$-1\leq x\leq 1$, this restricts our analysis of a spatially closed geometry 
($\Omega_k<0$) to redshifts lying in the interval 
$z\leq \sqrt{(\Omega_k -1)/\Omega_k}\, -1$. 
Note, however, that given the current estimates of the curvature parameter 
from WMAP and other experiments, i.e., $\Omega_k = -0.014 \pm {0.017}$%
~\cite{wmap}, the above interval leads to $z\leq 10$, which covers the entire 
range of current SNe Ia observations ($z \lesssim 2$). 

Now, by combining Eqs.~(\ref{dist-mod}), (\ref{dist-lumin}) and~(\ref{comoving-distance-WEC}), we obtain the following upper 
bound from the NEC/WEC:
\begin{widetext}
\begin{equation}  \label{WEC-bound}
\mu(z) \leq 5\,\log_{10}\,
\left[\frac{H_0^{-1}}{\sqrt{|\Omega_k|}}\,\,(1+z)\,S_{k}\left\{
S_{k}^{-1}\left[\sqrt{\left|\frac{\Omega_k}{\Omega_k-1}\right|}
\,\,(1+z)\right] -
S_{k}^{-1}\sqrt{\left|\frac{\Omega_k}{\Omega_k-1}\right|}\,\right\}
\,\right]+25 \;.
\end{equation}
\end{widetext}
Clearly, if the NEC/WEC are obeyed then $\mu(z)$ must take values such 
that Eq.~(\ref{WEC-bound}) holds. The three curves in top of Figure~(1a) 
show the NEC/WEC bounds on $\mu(z)$ as a function of the redshift for 
different signs of the curvature parameter $\Omega_k$. To plot the curves
we have used the central value of the \emph{HST} key project estimate for 
the Hubble parameter, i.e., $H_0 = 72$ $\rm{km \ s^{-1} Mpc^{-1}}$~\cite{hst} 
(that we assume throughout this paper). It is worth emphasizing that, as 
discussed in Ref.~\cite{SAR2006}, the predicted distance modulus depends 
very weakly on the value adopted for the Hubble parameter.

%%%%%%%%%%%%%%%%%%%%%%%%%%%
\vspace{0.3cm}
\centerline{\bf{B.  SEC}}
\vspace{0.3cm}
%%%%%%%%%%%%%%%%%%%%%%%%%%%

Similarly to the NEC, a first integration of Eq.~(\ref{sec-eq}) 
implies $\dot{a} \geq a_0  H_0$ $\forall$ $a< a_0$ which, along with 
Eqs.~(\ref{proper-distance}) and (\ref{I(a)}), gives the following 
upper bound on the radial comoving distance:
\begin{equation}  \label{comoving-distance-SEC}
r(z) \leq  \frac{H_0^{-1}}{a_0\sqrt{|\Omega_k|}}\,
S_{k}\left\{ \sqrt{|\Omega_k|}\,\ln(1+z) \right\}\;.
\end{equation}
Note that, differently from the previous case (NEC/WEC), the above 
constraint holds for any value of the curvature parameter $\Omega_k$. 
From Eqs.~(\ref{dist-mod}), (\ref{dist-lumin}) and (\ref{comoving-distance-SEC}),  
the SEC bound on $\mu(z)$ reads
\begin{eqnarray}   \label{SEC-bound}
\mu(z)  \leq  5\log_{10}\!\!\left[\!
\frac{(1+z)}{H_0\sqrt{|\Omega_k|}} %\times 
S_{k}\left\{ \sqrt{|\Omega_k|}\ln(1+z)
\right\} \!\right]\!\! +\! 25.\;\;
\end{eqnarray}
Figure~(1b) shows the SEC-$\mu(z)$ prediction as a function of the redshift. 
{}From top to bottom the curves correspond, respectively, to positive, null 
and negative values of the curvature parameter $\Omega_k$.

\begin{figure*}[t]
%\vspace{.2in}
\centerline{\psfig{figure=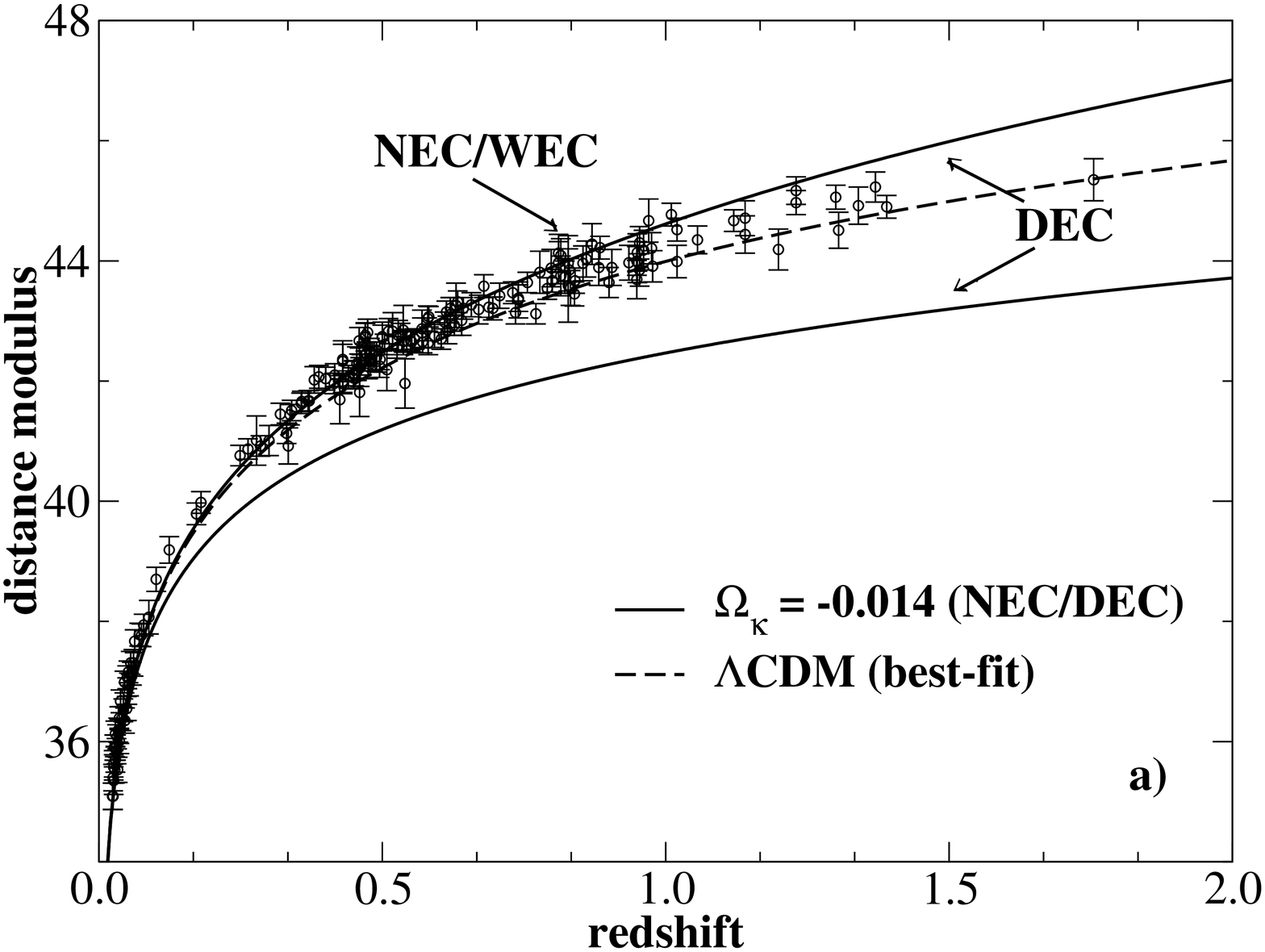,width=2.4truein,height=2.45truein,angle=-90} 
\psfig{figure=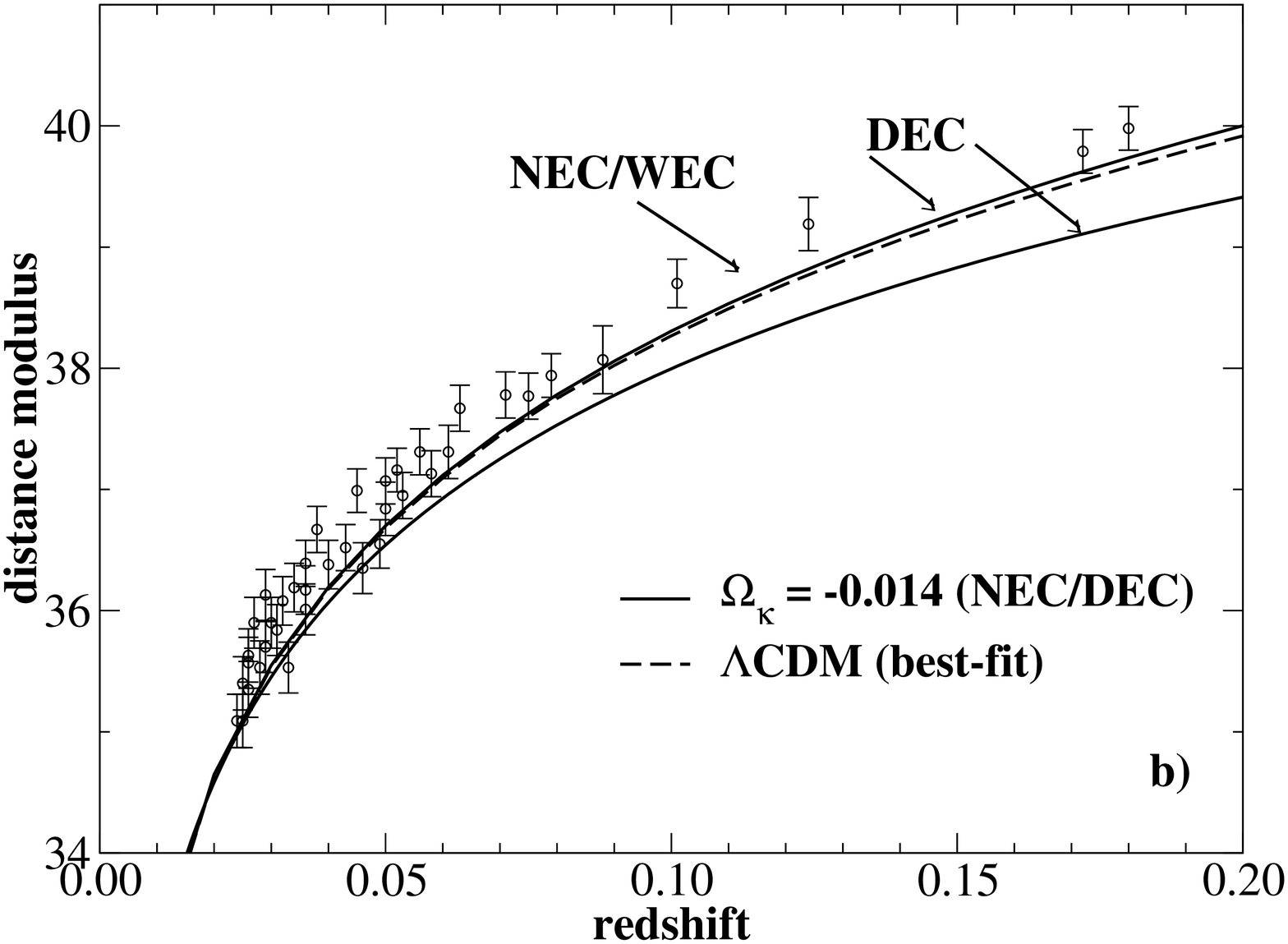,width=2.4truein,height=2.45truein,angle=-90}
\psfig{figure=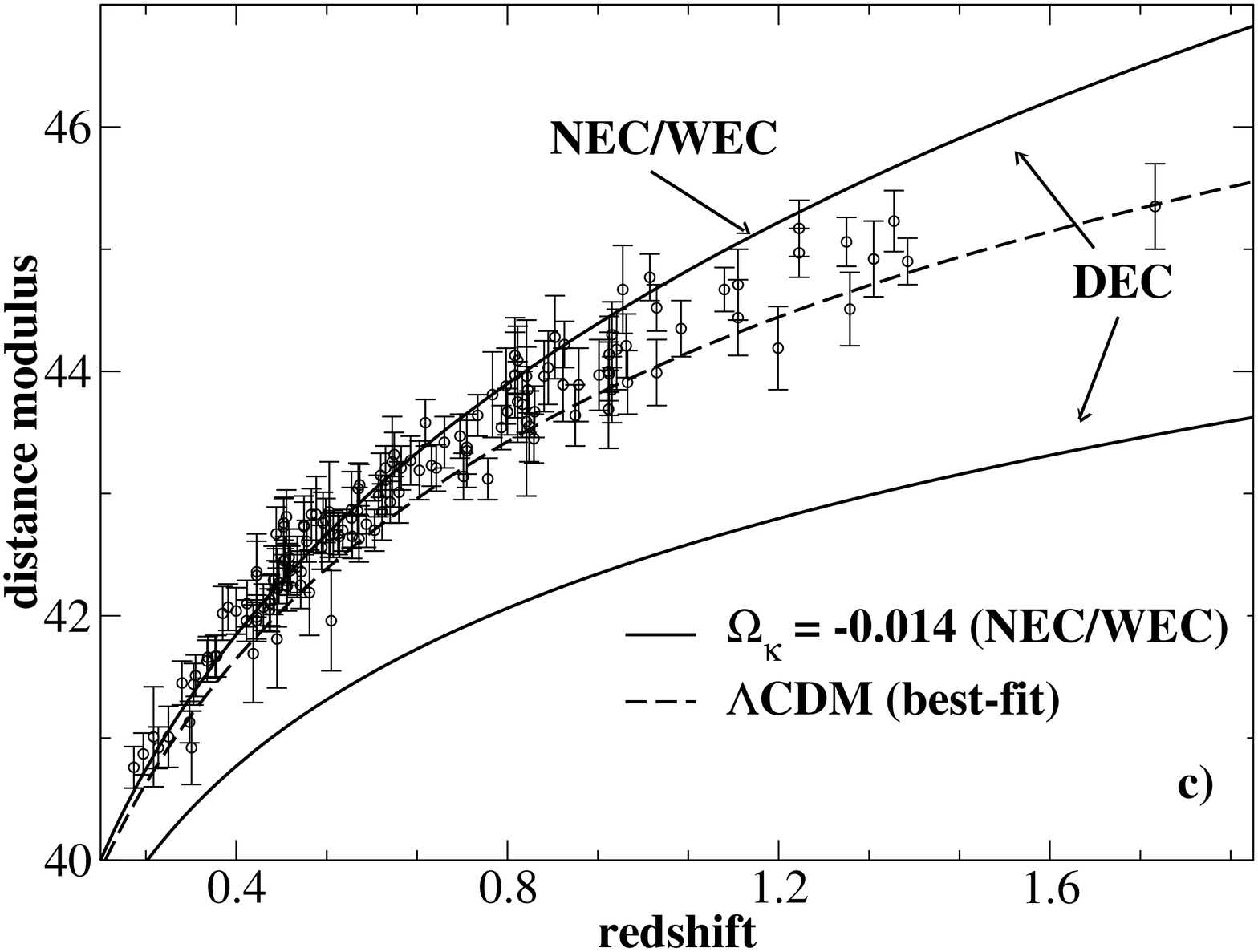,width=2.4truein,height=2.45truein,angle=-90}
\hskip 0.3in} 
\caption{NEC/WEC and DEC predictions for the distance modulus $\mu(z)$. The data points in the panels correspond to the new \emph{gold} sample of
182 SNe Ia. Panels {\bf (2a)} shows the  NEC/WEC and DEC bounds and the data points in the entire redshift interval $0.01 \lesssim z \lesssim 1.755$ while Panels {\bf (2b)} and {\bf (2c)} show the curves and data for a smaller range  of the redshift ($0.01 \lesssim z \lesssim 0.2$ and $0.2 \lesssim z \lesssim 1.755$, respectively.). As discussed in the text, these panels indicate that these energy conditions seem to have been violated by a considerable  number of nearby ($z \lesssim 0.2$) SNe Ia. For the sake of comparison, the best-fit $\Lambda$CDM model for the new \emph{gold} sample of 182 SNe Ia is also shown.} 
\end{figure*}

%\subsection{DEC}
%%%%%%%%%%%%%%%%%%%%%%%%%%%%%%%
\vspace{0.3cm}
\centerline{\bf{C.  DEC}}
\vspace{0.3cm}
%%%%%%%%%%%%%%%%%%%%%%%%%%%%%%%

DEC provides both upper and lower bounds on the rate of expansion. 
In order to find the lower bound from DEC we integrate the inequality 
on the left hand side of Eq.~(\ref{dec-eq}) to obtain 
$\dot{a}\leq\,a_0\,H_0\sqrt{\,\Omega_k + (1-\Omega_k)\,(a_0/a)^4}$. 
By combining this equation with Eqs.~(\ref{proper-distance}) and 
(\ref{I(a)}) we find
\begin{widetext}
\begin{eqnarray}
r(z) & \geq & \frac{H_0^{-1}}{ a_0\sqrt{|\Omega_k|}} \,
S_{k}\left\{-\frac{1}{2} S_{k}^{-1}\left[
\sqrt{\left|\frac{\Omega_k}{\Omega_k-1}\right|}\,
(1+z)^{-2}\right] \right. %\nonumber % \\
%& &
\left. +\frac{1}{2}
S_{k}^{-1}\sqrt{\left|\frac{\Omega_k}{\Omega_k-1}\right|}\,\right\}
  \label{comoving-distance-DEC}\,,
\end{eqnarray}
\end{widetext}
which holds for values of $\Omega_k <1$. Again, the above inequality, 
along with Eqs.~(\ref{dist-lumin}) and (\ref{dist-mod}), gives rise 
to the following lower bound on $\mu(z)\,$ from the DEC:
\begin{widetext}
\begin{equation}  \label{DEC-bound}
 \mu(z) \geq 5\,\log_{10}
\left[\frac{H_0^{-1}}{\sqrt{|\Omega_k|}}\,(1+z) \,
S_{k}\left\{\!-\frac{1}{2}
S_{k}^{-1}\left[\sqrt{\left|\frac{\Omega_k}{\Omega_k-1}\right|}\,
(1+z)^{-2}\right] +\frac{1}{2}\,
S_{k}^{-1}\sqrt{\left|\frac{\Omega_k}{\Omega_k-1}\right|}\,\right\}
\,\right]\!+ 25\;.
\end{equation}
\end{widetext}
As expected from Eq.~(\ref{dec-eq}), the DEC upper bound coincides with the NEC constraint on $\mu(z)$, which is given by Eq.~(\ref{WEC-bound}). Figure~(1a) illustrates this point, and also makes clear that the DEC-fulfillment gives rise to both a lower and an upper bounds on the distance modulus $\mu(z)$. It is also worth mentioning that, although we have restricted our analysis to the distance modulus, the inequalities (\ref{comoving-distance-WEC}), (\ref{comoving-distance-SEC}) and (\ref{comoving-distance-DEC}) are general bounds that can be used in any cosmological test involving the radial comoving distance.

\begin{figure*}[t]
%\vspace{.2in}
\centerline{\psfig{figure=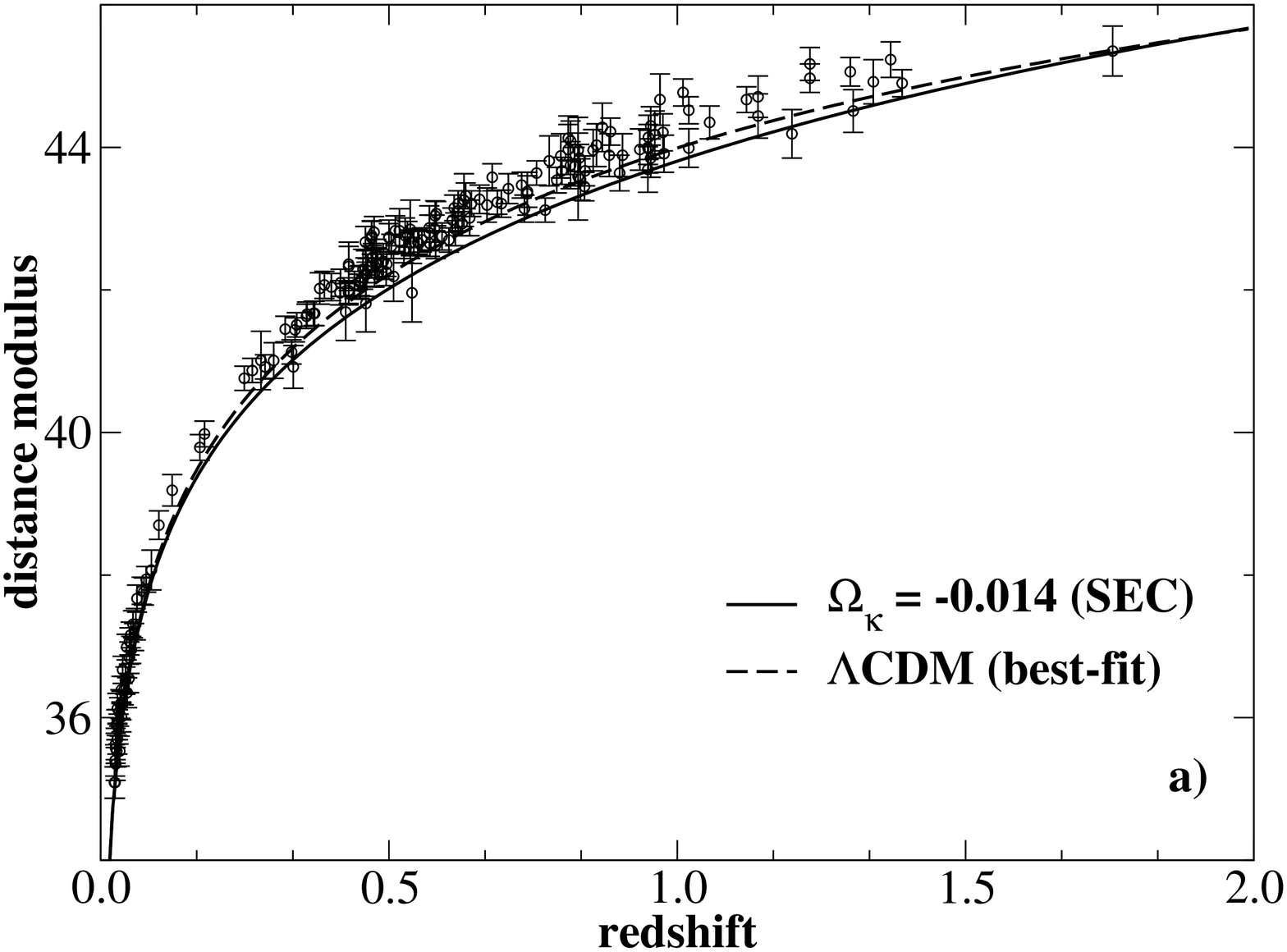,width=2.4truein,height=2.45truein,angle=-90} 
\psfig{figure=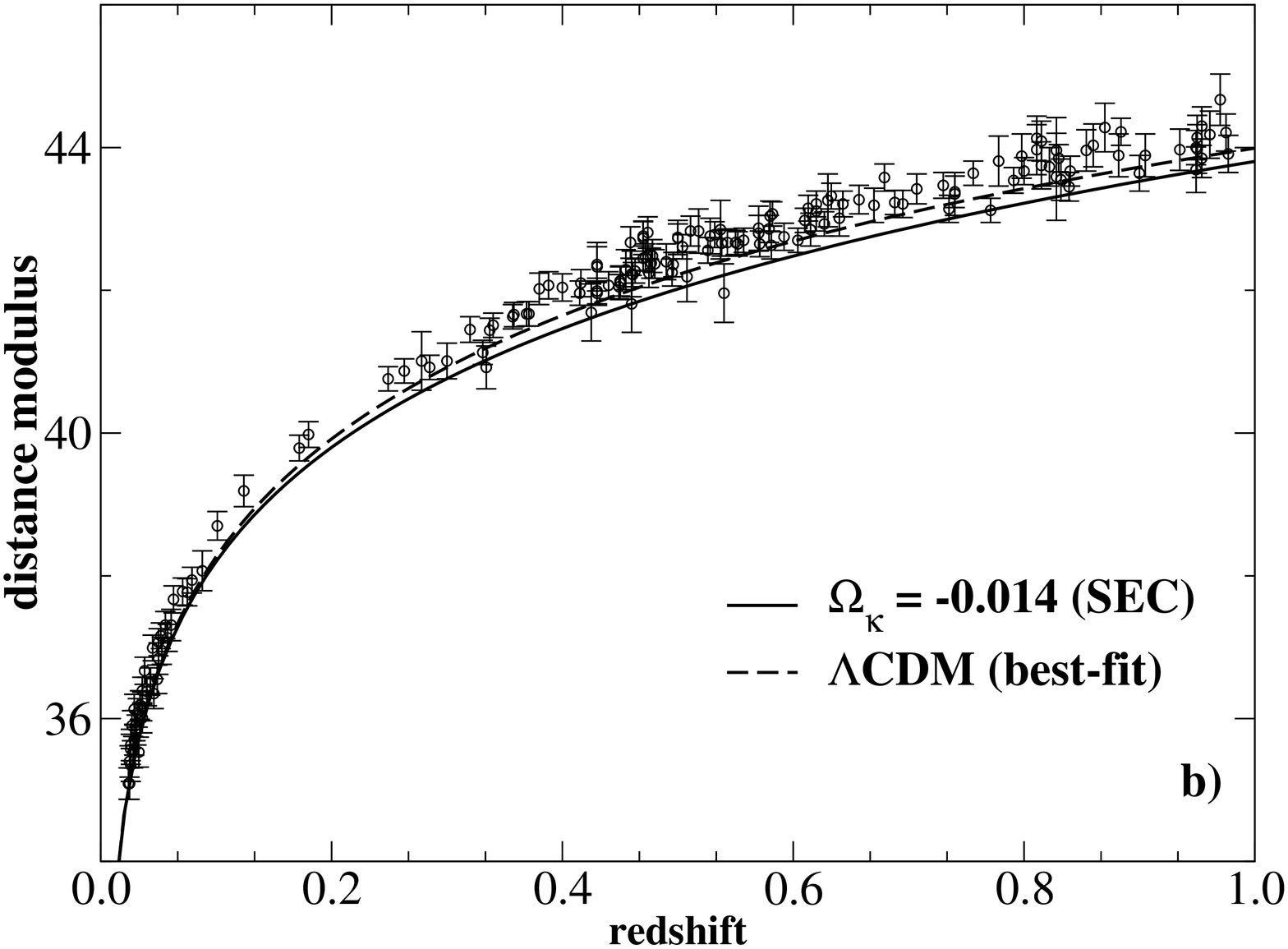,width=2.4truein,height=2.45truein,angle=-90}
\psfig{figure=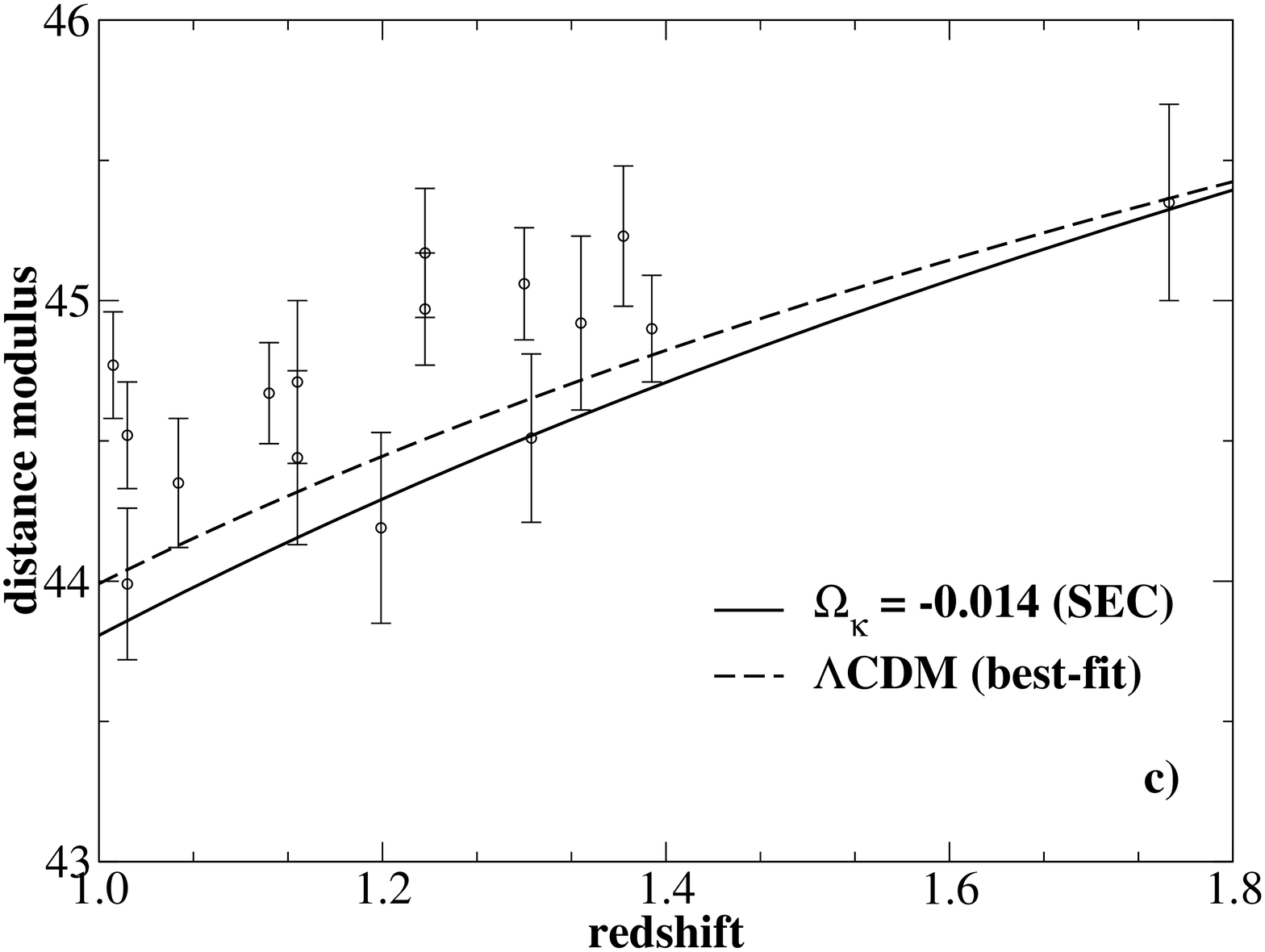,width=2.4truein,height=2.45truein,angle=-90}
\hskip 0.3in} 
\caption{The SEC upper bounds predictions for the distance modulus $\mu(z)$ are shown in the panels {\bf (3a)} to {\bf (3c)} for three  different redshift subintervals of $0.01 \lesssim z \lesssim 1.8$.  As in the previous Panels, the data points are from the new \emph{gold} sample of 182 SNe Ia. This figure shows that apart from a few SNe events the SEC seems to have been violated in the whole redshift range. As discussed in the text, the panel {\bf (3c)} shows that even at hight redshifts  ($z \gtrsim 1$) a considerable number of SNe Ia points lie above the SEC-fulfillment curves. In all Panels the best-fit $\Lambda$CDM model for the new \emph{gold} sample of 182 SNe Ia is also shown.} 
\end{figure*}

%%%%%%%%%%%%%%%%%%%%%%%%
\section{Results and Discussion}
%%%%%%%%%%%%%%%%%%%%%%%%

In Figs.~(2) and~(3) we confront the energy conditions predictions for $\mu(z)$ 
with current SNe Ia observations. The data points appearing in the panels 
correspond to the new \emph{gold} sample of 182 events distributed over the 
redshift interval $0.01 \lesssim z \lesssim 1.7$, as compiled by 
Riess \emph{et al.} in Ref.~\cite{Riess2006}, which include the new, recently 
discovered 21 SNe Ia by the \emph{HST}. In order to perform our subsequent 
analyses, from now on  we adopt  $\Omega_k = -0.014$, which corresponds to 
the central value of the estimates provided by current WMAP  
experiments~\cite{wmap}. For the sake of comparison, the best-fit $\Lambda$CDM model for the new \emph{gold} sample (corresponding to $\Omega_m \simeq 0.48$ and $\Omega_{\Lambda} \simeq 0.95$) is shown in all Panels of Figs. (2) and (3).

Figures~2(a)--~2(c) show the upper and lower-bound curves $\mu(z)$ for the 
NEC/WEC and DEC-fulfillment. As discussed in Ref.~\cite{SAR2006},  an interesting 
aspect of these panels is that they indicate that these energy conditions 
might have been violated by a considerable number of nearby SNe Ia, at 
$z \lesssim 0.2$. To better visualize that we show a closer look of the data 
points in this interval (Panel 2b) and take as example the cases of the SNe 
1992bs, 1992aq and 1996ab which are, respectively, at $z = 0.063$, $z = 0.101$ 
and $0.124$. While their observed distance modulus are 
$\mu_{\rm{1992bs}} =  37.67 \pm 0.19$, $\mu_{\rm{1992aq}} =  38.70 \pm   0.20$, 
and  $\mu_{\rm{1996ab}} =  39.19 \pm 0.22$, the upper-bound NEC predictions 
for the corresponding redshifts gives, respectively, $\mu(z = 0.063) = 37.22$, 
$\mu(z = 0.101) = 38.33$ and $\mu(z = 0.124) = 38.82$. In the case of the SN 1992bs, 
for instance, we note that the discrepancy between the observed value and the NEC/WEC 
prediction is of $0.45$ in magnitude or, equivalently, $\simeq 2.36 \sigma$, which 
clearly indicates a violation of NEC/WEC at this redshift. The largest discrepancy, 
however, is associated with the observations of the SN 1999ef at $z = 0.038$ 
and $\mu_{\rm{1999ef}} = 36.67 \pm 0.19$.  In this case, the upper-bound NEC/WEC 
prediction is  $\mu(z = 0.038) = 36.08$, which is $\simeq 3.15$ off from the 
central value measured by the High-$z$ Supernovae Team \cite{Riess2006}.

Concerning the above analysis it is also worth emphasizing three important 
aspects at this point.
First,  that the above results holds for the upper-bound DEC 
predictions, and that the lower-bound of DEC is not violated by the current
SNe Ia data. Also, neither NEC/WEC nor DEC are violated at higher redshifts, 
i.e., at $z > 0.2$ (Fig. 2c). 
Second, the analysis is very insensitive to the values of the curvature 
parameter in that all the above conclusions are unchanged for values 
of $\Omega_k$ lying in the interval provided by the current CMB experiments, 
i.e., $\Omega_k = -0.014 \pm {0.017}$~\cite{wmap}.%
\footnote{For instance, by taking the upper and lower 1$\sigma$ limit given 
by WMAP, i.e., $-0.031 \leq \Omega_k \leq 0.003$, the NEC/WEC predicted 
distance modulus at $z = 0.038$ ranges between $\mu(z = 0.038) = 36.081$ 
and $36.079$, respectively.}
Finally, we note that, although our analyses and results are completely 
model-independent, in the context of a FLRW model with a dark energy 
component parameterized by an equation of  state (EoS) $w \equiv p/\rho$, 
violation of NEC/WEC and DEC is associated with the existence of the 
so-called \emph{phantom} fields ($w < -1$), an idea that has been 
largely explored in the current literature~\cite{caldwell}. 
By assuming this standard framework, the results above, therefore, 
seem to indicate a possible dominion of these fields over the 
conventional matter fields very recently, at $z \lesssim 0.2$. 

SEC is the requirement (easily verified in our everyday experience) that 
gravity is always attractive. In an expanding FLRW universe, this amounts 
to saying that cosmic expansion must be decelerating regardless of the 
sign of the spatial curvature, as mathematically expressed by 
Eq.~(\ref{sec-eq}).%
\footnote{As is well knwon, an early period of cosmic deceleration is 
strongly supported by the  primordial nucleosynthesis predictions and 
the success of the gravitational instability theory of structure formation.} 
Similarly to the NEC/WEC/DEC analysis, one can also estimate the epoch of the first SEC violation by mapping the current SNe Ia Hubble diagram.

The upper-bound curves $\mu(z)$ for the SEC-fulfillment are shown in 
Figs.~(3a)--(3c) for three different redshift intervals. Note that 
SEC seems to be violated in almost the entire redshift range, with only 
very few SNe events in agreement with the theoretical upper-bound SEC 
prediction. In particular, we note that even at very high redshifts, i.e., 
$z \gtrsim 1$, when the Universe is expected to be dominated by normal matter, 
eleven out of sixteen SNe Ia measurements provide magnitude at least $1\sigma$ 
higher than the theoretical value derived from Eq.~(\ref{SEC-bound}). 
As an example, let us consider the cases of the SNe HST04Sas (the highest-$z$ 
SN to violate SEC) at $z = 1.39$ and $\mu_{\rm{HST04Sas}} = 44.90 \pm 0.19$ 
and  HST05Koe at $z = 1.23$ and $\mu_{\rm{HST05Koe}} = 45.17 \pm 0.23$. 
While the distance modulus of the former is at the limit of $\simeq 1\sigma$ 
higher than the SEC prediction [$\mu(z = 1.39) = 44.68$], the observed value 
of $\mu(z)$ for the latter is $\simeq 3.5\sigma$ far from the theoretical 
value of Eq.~(\ref{SEC-bound}) [$\mu(z = 1.23) = 44.36$], a discrepancy that 
clearly indicates violation of SEC at $z > 1$. An interesting aspect worth 
mentioning is that if the redshift of these first SNe Ia events that violate 
SEC could be taken as the beginning of the epoch of cosmic acceleration ($z_a$), 
then our current concordance scenario (a flat $\Lambda$CDM model with 
$\Omega_m \simeq 0.3$), whose prediction is $z_a \simeq 0.67$, would be in 
disagreement with this estimate. In reality,  for the current accepted 
dark matter-dark energy density parameter ratio (of the order of 
${\Omega_{m}}/{\Omega_{x}} \sim 0.4$), the entire class of flat models 
with a constant EoS $w$, which predicts 
$z_a = [-(3w+1)\frac{\Omega_{x}}{\Omega_m}]^{-1/3w} - 1$, 
also would be in disagreement with the first redshifts of SEC violation 
discussed above.

\section{Concluding Remarks}

In this work, by extending and updating previous results~\cite{SAR2006}, we have derived, from  the classical energy conditions, model-independent bounds on the behavior of the distance  modulus of extragalactic sources as a function of the redshift for any spatial curvature. We have also confronted these  energy-condition-fulfillment bounds with the new \emph{gold} sample of 182 SNe observed events provided  recently by Riess \emph{et al.\/} in Ref.~\cite{Riess2006}. On general grounds, our analyses indicate that the NEC/WEC and DEC are violated by a significant number of low-$z$ ($z \lesssim 0.2$) SNe Ia, while for higher redshifts none of these energy conditions  seems to have been violated. Another important outcome of our analyses is that the SEC, whose violation in expanding FLRW model is ultimately related to the  cosmic acceleration, seems to be violated in the entire  redshift range covered by the new SNe Ia \emph{gold} sample. A surprising fact from the confrontation between the SEC prediction  and  SNe Ia observations is that this energy condition seem to have been firstly violated at very high-$z$ ($\simeq 1.3$), which is very far from the transition redshift predicted by most of the quintessence models and by the current standard  concordance flat $\Lambda$CDM  scenario ($z \simeq 0.67$). In agreement with our previous analysis~\cite{SAR2006}, we  emphasize that  the results reported here reinforce the idea that, in the context of the standard cosmology, no possible combination of \emph{normal} matter is capable of fitting the current observational data.

\begin{acknowledgments}
JS and NP acknowledges the support of PRONEX (CNPq/FAPERN).  JSA and MJR thank CNPq for the grants under which this work was carried out.  JSA is also supported by Funda\c{c}\~ao de Amparo \`a Pesquisa do Estado do  Rio de Janeiro (FAPERJ) No. E-26/171.251/2004.
\end{acknowledgments}

\end{document}